\documentclass{ws-mpla}
\usepackage[super]{cite}
\usepackage{graphicx}
\usepackage{color}
\begin{document}

\markboth{Y.Li and L.Y.Zhang}{Inflationary magnetogenesis with a self-consistent coupling function}

\catchline{}{}{}{}{}

\title{Inflationary magnetogenesis with a self-consistent coupling function
}

\author{Yu Li\footnote{
Corresponding author.}}

\address{School of Science, Dalian Maritime University, Dalian 116026, China\\
leeyu@dlmu.edu.cn}

\author{Le-Yao Zhang}

\address{School of Science, Dalian Maritime University, Dalian 116026, China\\
	zhangleyao@dlmu.edu.cn}

\maketitle

\pub{Received (Day Month Year)}{Revised (Day Month Year)}

\begin{abstract}
In this paper, we discuss the inflationary magnetogenesis scenario, in which the coupling function is introduced to break the conformal invariance of electromagnetic action. Unlike in conventional models, we
deduce the Maxwell's equations under the perturbed FRW metric.  
We found that, the self-consistency of the  action depends on the form of the coupling function when the scalar mode perturbations have been considered. Therefore, this self-consistency can be seen as a restriction on the coupling function. In this paper, we give the restrictive equation for coupling function then obtain the specific form of the coupling function in a simple model. We found that the coupling function depends on the potential of the inflaton and thus is model dependent. We obtain the 
power spectrum of electric field and magnetic field in large-field inflation model. 
We also found that the coupling function is a incresing function of time during slow-roll era as 
most of inflationary magnetogenesis models, it will lead to strong coupling problem. 
This issue is discussed qualitatively by introducing a correction function during the preheating.

\keywords{Primordial magnetic field; inflationary magnetogenesis; cosmology perturbations.}
\end{abstract}

\ccode{PACS Nos.:98.80.Cq.}

\section{Introduction}
Observations indicate that the universe is magnetized on a wide range of length scales\cite{r2,r3,r4,r5,r6,r7,r8}.
The sources of these magnetic fields are still unclear. 
There are two types of models that can be used to explain the origin of these magnetic fields: astrophysical scenario
\cite{r9,r10,r11,r12} and primordial scenario(see Refs.\cite{r1,r36,r38,r39} for reviews). The former believes that these magnetic fields originate from some astrophysical processes. 
The origin of magnetic fields in galaxies and clusters  can be explained in such models. However, 
this type of models is difficult to explain the origin of the magnetic fields in cosmic voids. 
The magnetic fields in the cosmic voids are more like the origin of the early universe\cite{r6,Archambault_2017,Ackermann_2018}. 
The latter, i.e. primordia scenario, assumes that these large-scale magnetic fields originated in the early stage of the universe.

One class of possible sources of the primordial magnetic fields are phase transitions like electroweak phase transition \cite{r14,r16,Vachaspati_2021} or the QCD transtition \cite{r17,r21}. 
However, in these scenarios very tiny fields on galactic scales obtain unless helicity is also generated, in which case one can have an inverse cascade of energy to large scales\cite{r52,r53}.
The other class of possible sources of primordial magnetic field are inflationary magnetogenesis\cite{r13,r15,r22,r28,r29,r32,r33,r40,r41,PhysRevD.100.023524,Fujita_2019}. The inflation provides an ideal setting for the generation of primordial large-scale field\cite{r13},
therefore we focus on the inflationary magnetogensis in this paper.

Due to the conformal invariance of the standard electromagnetic action and the FRW metric is conformally flat, the electromagnetic field is not amplified during the inflation era \cite{r42}. Therefore, in order to be able to generate large scale magnetic fields by inflation, it is necessary to break this conformal invariance\cite{r13,r15,r22,r48,r49,r50,r51,r37}.
One way to do this is to introduce a time-dependent coupling function $f^2(\phi)$ into
the action \cite{r15}. 

On the other hand, an effective way of linking theoretical models to observations is to consider the effects of the existence of large-scale magnetic fields on cosmological perturbations.
While, cosmological perturbations will also in turn affect the evolution of large-scale magnetic fields. So the complete discussion should be to solve the evolution of electromagnetic field and cosmological perturbation together. In other words, it is necessary to consider the magentogenesis in inflation model in which cosmological perturbations have been included.

Hovever, it is difficult to solve the equations which include all fields (pertubations and electromagnetic field). There are two methods to discuss this issue approximately. One way is to consider the magnetogenesis in unperturbed FRW metric and discuss its backreaction on the perturbations e.g. on CMB. Most of the current work is done in this way (see \cite{r54,r55,r56,PhysRevD.101.103526} for example). The other way, which is used in this paper, is to consider the magnetogenesis in perturbed FRW metric and discuss the influence of perturbations on the electromagnetic field.

As we will discuss in this paper, the existence of cosmological perturbations restrict the form of the coupling function.
Under the FRW background, the introduction of the coupling function does not destroy the self-consistency of the action, which means that the secondary constraint equation for electromagnetic field $\vec{\nabla}\cdot\vec{E}=0$ is satisfied automatically. However, if one consider the perturbed FRW background, this constraint equation will be not a trivial equation. In this situation, we can treat this equation as the restriction on the coupling function, i.e. the form of a self-consistent coupling function $f(\phi)$ should satisfy this equation. The purpose of this paper is to discuss the inflationary magnetogensis with this self-consistent coupling function.

This paper is organized as follow: We deduce the Maxwell's equations under the perturbed FRW metric , and then get the restrict equation for $f(\phi)$ in section \ref{s2}~. We apply this restrict equation to 
slow-roll inflation in the end of section \ref{s2}~ and obtain the power spectrum of electromagnetic field in large-field inflation model in section \ref{s3}~. We also discuss the backreaction in section \ref{s3}~ and strong coupling problems in section \ref{s4}~, the summary in the section \ref{s5}~.
\section{\label{s2}Maxwell's equations under perturbated FRW background}

To get the restrict equation for $f(\phi)$, let us consider the FRW metric with scalar mode of inhomogeneous perturbations in longitudinal gauge:
\begin{eqnarray}
	ds^2&=&-(1+2\Phi)dt^2+a^2(t)(1-2\Phi)\delta_{ij}dx^idx^j\label{e1}\\
	&=&a^2(\eta)\left[-(1+2\Phi)d\eta^2+(1-2\Phi)\delta_{ij}dx^idx^j\right]\label{e2}
\end{eqnarray}
where $\Phi$  is Bardeen potential, $t$  is cosmic time and $\eta$  is conformal time.
The action of matter during inflation can be written as \cite{r15}:
\begin{equation}
	S=-\frac{1}{16\pi}\int d^4x \sqrt{-g}\left[g^{\alpha\beta}g^{\mu\nu}f^2(\varphi)F_{\mu\alpha}F_{\nu\beta}\right]
	-\int d^4x \sqrt{-g}\left[\frac{1}{2}g^{\mu\nu}\partial_{\mu}\varphi\partial_{\mu}\varphi+V(\varphi)\right]\label{e3}
\end{equation}
where $\varphi(t,{\bf x})=\phi(t)+\delta\phi(t,{\bf x})$ is the inflaton and its perturbation. 
$F_{\alpha\beta}=A_{\beta;\alpha}-A_{\alpha;\beta}=A_{\beta,\alpha}-A_{\alpha,\beta}$
is the electromagnetic field tensor, with $A_{\alpha}$ being the standard electromagnetic 4-potential. $f(\varphi)$ is the coupling function which is introduced to break the conformally invariant of
the standard electromagnetic action \cite{r15}. For the convenience of discussion, we expand the coupling function as:
\begin{equation}
	f^2(\varphi)=f^2(\phi+\delta\phi)\approx\left[f(\phi)+\frac{df}{d\varphi}\Big|_{\phi}\delta\phi\right]^2
	\approx f^2(\phi)\left[1+\mathcal{G}(\phi)\delta\phi\right]\label{e6}
\end{equation}
where
\begin{equation}\label{e7}
	\mathcal{G}(\phi)\equiv\frac{2}{f(\phi)}\frac{df}{d\varphi}\Big|_{\phi}
\end{equation}
It is worth to notice that $\mathcal{G}$ depend only on time or, in other words, it is scale-independent.

In the model we discuss here, we treat the electromagnetic field as a ``test" field which means
that $F_{\alpha\beta}$  do not affect the evolution of the background ($a$ and $\phi$) and perturbations ($\Phi$ and $\delta\phi$),
but the background and perturbations can affect the evolution of electromagnetic field.

The Maxwell's equations can be obtain by using the action (\ref{e3}):
\begin{equation}\label{e4}
	\partial_{\rho}\left[\sqrt{-g}f^2(\varphi)g^{\sigma\mu}g^{\rho\nu}F_{\mu\nu}\right]=0
\end{equation}
For conformal time, the time component  of Eq.(\ref{e4}) ($\sigma=0$) lead to:
\begin{equation}\label{e8}
	\partial_i\left[f^2(\phi)(1-2\Phi+\mathcal{G}\delta\phi)\delta^{ij}F_{0j}\right]=0
\end{equation}
In Minkowski spacetime, Eq.(\ref{e8}) is noting but $\vec{\nabla}\cdot\vec{E}=0$.
This is the secondary constraint equation for source-free  electromagnetic field (one can refer to Appendix E in \cite{b1}).We will see later, this equation is trivial equation in FRW background and is non-trivial in perturbed background.

The space component of Eq.(\ref{e4}) ($\sigma=i$) can also be obtain similarly:
\begin{eqnarray}
	&&(1-2\Phi+\mathcal{G}\delta\phi)A''_j-(1+2\Phi+\mathcal{G}\delta\phi)\nabla^2A_j\nonumber\\
	&&+\left[\mathcal{G}\phi'(1-2\Phi+\mathcal{G}\delta\phi)-2\Phi'+\mathcal{G}'\delta\phi+g\delta\phi'\right]A'_j\nonumber\\
	&&+(2\Phi_{,k}+\mathcal{G}\delta\phi_{,k})
	\delta^{k\ell}(A_{\ell,j}-A_{j,\ell})=0\label{e9}
\end{eqnarray}
where $\nabla^2\equiv\delta^{k\ell}\partial_k\partial_{\ell}$ is Laplace operator and $'$ denote the derivative with respect to conformal time.
For the convenience of discussion, we assume that $A_i$ can be expressed as perturbation expansion:
\begin{equation}\label{e13}
	A_i=A_i^{(0)}+A_i^{(1)}+A^{(2)}_i+\cdots
\end{equation}
where $O[A_i^{(0)}]\sim O[\Phi]$.

In this paper we adopt the Coulomb gauge:$A_0(\eta,{\bf x})=0,~\partial_jA^j(\eta,{\bf x})=0$. 
Under this gauge, Maxwell's equations can be rewritten as:
\begin{equation}\label{e21}
	\partial_i\left\{\mathcal{G}\delta\phi[A^{(0)}_j]'-4\Phi[A^{(0)}_j]'-2A^{(0)}_j\Phi'\right\}\delta^{ij}=0
\end{equation}
and
\begin{equation}
	\left\{[A^{(0)}_j]''+\mathcal{G}\phi'[A^{(0)}_j]'-\nabla^2A^{(0)}_j\right\} 
	+ \left\{[A^{(1)}_j]''+\mathcal{G}\phi'[A^{(1)}_j]'-\nabla^2A^{(1)}_j\right\}-Q_j^{(1)}=0\label{e22}
\end{equation}
where:
\begin{eqnarray}
	Q_j^{(1)} &\equiv&(2\Phi-\mathcal{G}\delta\phi)  \left\{[A^{(0)}_j]''+\mathcal{G}\phi'[A^{(0)}_j]'-\nabla^2A^{(0)}_j\right\} \nonumber\\
	&+&(2\Phi-\mathcal{G}\delta\phi)'[A^{(0)}_j]'
	+4\Phi\nabla^2A^{(0)}_j+\delta^{k\ell}(\mathcal{G}\delta\phi+2\Phi)_{,k}(A^{(0)}_{j,\ell}-A^{(0)}_{\ell,j})\label{e23}
\end{eqnarray}
Notice that, all the order in the left of Eq.(\ref{e22}) should be zero, therefore the space component of Maxwell equations lead to
\begin{eqnarray}
	[A^{(0)}_j]''+\mathcal{G}\phi'[A^{(0)}_j]'-\nabla^2A^{(0)}_j&=&0\label{e24}\\
	~[A^{(1)}_j]''+\mathcal{G}\phi'[A^{(1)}_j]'-\nabla^2A^{(1)}_j&=&Q_j^{(1)}\label{e25}
\end{eqnarray}

If scalar perturbations are not taken into account ($\Phi=\delta\phi=0$), then time component of Maxwell's equations Eq.(\ref{e21}) becomes a trivial equation, and source term $Q_j^{(1)}$ vanish. One can 
choose the form of the coupling function $f(\phi)$ (and $\mathcal{G}$) freely from the beginning and solve the space component Eq.(\ref{e24},\ref{e25}) directly. However, once the scalar perturbations have been considered, Eq.(\ref{e21}) is no longer trivial. 
This means that not any coupling function can make the theory self-consistent.
The form of the coupling function is to ensure that this constraint equation holds.
Although Eq.(\ref{e21}) is order $O[\Phi^2]$, it does not means that the restriction on coupling function is order of perturbations. The trivial or non-trivial of Eq.(\ref{e21}) is essentially different.
This fact is not surprising. 
It is rooted in the fact that the choice of Lagrangian density is not arbitrary and it needs to satisfy self-consistent conditions\cite{b2}.

From Eq.(\ref{e21},\ref{e24},\ref{e25}) one can see that, there are two aspects to the influence of cosmological perturbations on electromagnetic fields: 
\begin{enumerate}
	\item The perturbations restrict the form of coupling function by Eq.(\ref{e21}).
	\item The perturbations (and $A^{(0)}_j$)  provide a source term $Q_j^{(1)}$ for $A^{(1)}_j$.
\end{enumerate}
An consistent discussion should be solving Eq.(\ref{e21},\ref{e24},\ref{e25}) together. However, one can notice that, there is no $A^{(1)}_j$ in Eq.(\ref{e21}) and Eq.(\ref{e24}). This means that we can solve Eq.(\ref{e21},\ref{e24}) to obtain the $\mathcal{G}$ and $A^{(0)}_j$ first, and then insert the $\mathcal{G}$ and $A^{(0)}_j$ into Eq.(\ref{e25}) to get the solution of $A^{(1)}_j$.

Furthermore, the source term $Q_j^{(1)}$ in Eq.(\ref{e25}) is order $O[\Phi^2]$, and then the $A^{(1)}_j$ will be more smaller than $A^{(0)}_j$. Therefore in this paper, we only focus on the evolution of the main part of $A_j$, i.e. $A_j^{(0)}$. In other words, we only focus on the influence (i) of 
cosmological perturbations. 

In fact, Eq.(\ref{e24}) is the same as the evolution equation of the electromagnetic 4-potential in the conventional model \cite{r37} besides the coupling function and $A^{(0)}_j$ should satisfy the Eq.(\ref{e21}). 
We  treat Eq.(\ref{e21}) as a restriction on the form of the coupling function. 
One can choose a coupling function which satisfy Eq.(\ref{e21}), then to solve the Eq.(\ref{e24}).

\textcolor{black}{
	From Eq.\eqref{e21}, one can obtain that  
	\begin{equation}
		\label{e26}
		\mathcal{G}\delta\phi[A^{(0)}_j]'-4\Phi[A^{(0)}_j]'-2A^{(0)}_j\Phi'=\mathbb{C}(\eta)
	\end{equation}
where $\mathbb{C}(\eta)$ is the function of $\eta$. The choice of function $\mathbb{C}(\eta)$ will affect the form of the coupling function.
}

%
\textcolor{black}{
In order to get the specific form of the coupling function, we consider the slow-roll era of inflation first.
During the slow-roll inflation, the super-Hubble scale Fourier mode of Bardeen potential $\Phi_k$ satisfy\cite{r44,r45}
\begin{equation}
	\Phi'_k = 0,~~~~
	\Phi_k\approx\epsilon\mathcal{H}\frac{\delta\phi_k}{\phi'}\label{e27}
\end{equation}
where $\epsilon\equiv-\dot{H}/H^2$ is slow-roll parameter,  and dot denote the derivative with respect to cosmic time. $\mathcal{H}\equiv a'/a$
is conformal Hubble parameter and $\delta\phi_k$ is Fourier mode of $\delta\phi$.
If one consider the large scales only, then we have:
\begin{equation}
	\Phi' \approx 0,~~~~
	\Phi\approx\epsilon\mathcal{H}\frac{\delta\phi}{\phi'}\label{e30}
\end{equation}
Insert Eq.(\ref{e30}) into Eq.(\ref{e26}) one can get
\begin{equation}\label{e31-1}
	\mathcal{G}=4\frac{\epsilon\mathcal{H}}{\phi'}+\frac{\mathbb{C}(\eta)}{\delta\phi[A^{(0)}_j]'}
\end{equation}
From the Eq.\eqref{e7}, one can see that  $\mathcal{G}$ is scale-independent, therefore the second term in righthand of Eq.\eqref{e31-1} should vanish, which means that we should consider the models in which $\mathbb{C}(\eta)=0$. This consideration give
\begin{equation}\label{e31}
	\mathcal{G}=4\frac{\epsilon\mathcal{H}}{\phi'}
\end{equation}
}

Although the above discussion only considers the behavior at large scales, the form of the $\mathcal{G}$ is scale-independent, so Eq.(\ref{e31}) holds at all scales.

During the slow-roll inflation, the Klein-Gorden equation of $\phi$ is $3H\dot{\phi}\approx-V_{,\phi}$. Note that the Friedmann equation in the slow-roll 
inflation is $H^2\approx V/3$, then Eq.(\ref{e31}) change to 
\begin{equation}\label{e32}
	\mathcal{G}=-2\frac{V_{,\phi}}{V}
\end{equation}
Using Eq.(\ref{e7}) one can get the form of coupling function $f(\phi)$ as:
\begin{equation}\label{e33}
	f(\phi)\propto\exp\left(-\int^{\phi}\frac{V_{,\phi}}{V}d\phi\right)\propto V^{-1}
\end{equation}
Eq.(\ref{e33}) shows that the form of self-consistent coupling functiong depend on the 
potential of inflaton. In other words, the form of the coupling function is model dependent.

\section{\label{s3}Power spectrum of electromagnetic field in large-field model}
To obtain the power spectrum of electromagnetic field, we consider the 
large-field inflation with polynomial potentials:
\begin{equation}
	\label{e38}
	V(\phi)=\Lambda^4\left(\frac{\phi}{\mu}\right)^p~~~~(p>0)
\end{equation}
where $\Lambda$ is the ``height" of potential, corresponding to the vacuum energy density during inflation, and $\mu$ is the ``width" of the potential, corresponding to the change in the field value $\Delta\phi$ during inflation\cite{r45}.

According to Eq.(\ref{e33}), the coupling function in this large-field model is 
\begin{equation}
	\label{e39}
	f(\phi)\propto\phi^{-p}
\end{equation}

By using $\dot{\phi}\approx-V,_{\phi}/3H,~~3H^2\approx V$ during slow-roll inflation one can have\cite{b3}:
\begin{equation}
	\ln \frac{a}{a_i}=-\frac{1}{2p}(\phi^2-\phi_i^2)\label{e41}
\end{equation}
where $a_i$ and $\phi_i$ is the scale factor and value of inflaton at the beginning of the inflation. And then the coupling function can be written as the function of scale factor:
\begin{equation}
	\label{e42}
	f(a)\propto\left(-2p\ln \frac{a}{a_i}+\phi_i^2\right)^{-p/2}
\end{equation}
This form of $f(\phi)$ is different from conventional models (see \cite{r39} for review). In these models, it is often assumed that the coupling function has a power law form of scale factor. 
However, Eq.(\ref{e42}) shows that the form of the self-consistent coupling function is not a power law form as in conventional models. 
This means that the power-law-like coupling function does not 
satisfy the self-consistent condition Eq.(\ref{e26}). This is the main difference between the model we discuss here and the conventional models.

During slow-roll inflation era, the scale factor as the function of confomal time $a(\eta)$ can be assumed
as:
\begin{equation}
	\label{e43}
	a(\eta)=a_i\left|\frac{\eta}{\eta_i}\right|^{1+\beta}
\end{equation}
where $\eta_i$ is the conformal time when the inflation begin. The case $\beta=-2$ corresponds to de Sitter space-time. During the inflation, $\eta\rightarrow0_{-}$. Insert Eq.(\ref{e43})  into Eq.(\ref{e42}) we have:
\begin{equation}
	\label{e44}
	f\propto\left[-2p(1+\beta)\ln\left|\frac{\eta}{\eta_i}\right|+\phi^2_i\right]^{-p/2}
\end{equation}

It is worth noting that the coupling function will diverges at the conformal time
\begin{equation}
	|\eta_{\infty}|=|\eta_i|\exp\left[\frac{\phi^2_i}{2p(1+\beta)}\right]\label{e46}
\end{equation}
When $\eta=\eta_{\infty}$, $\phi=0$, which means that the slow roll phase has been ended before $\eta_{\infty}$.

Insert Eq.(\ref{e46}) into Eq.(\ref{e44}) one can get:
\begin{equation}
	\label{e47}
	f\propto\left[2p(1+\beta)\ln\left|\frac{\eta_{\infty}}{\eta}\right|\right]^{-p/2}\propto\left[\ln\left|\frac{\eta_{\infty}}{\eta}\right|\right]^{-p/2}
\end{equation}

The next step is to solve Eq.(\ref{e24}) by using coupling function Eq.(\ref{e44}) or Eq.(\ref{e47}). Before that, it is convenient to 
set $\mathcal{A}\equiv a(\eta)f(\phi(\eta))A^{(0)}_j(\eta,k)$, where $A^{(0)}_j(\eta,k)$
is Fourier mode of $A^{(0)}_j$. Eq.(\ref{e24}) can be written as\cite{r39}:
\begin{equation}\label{e35}
	\mathcal{A}''(\eta,k)+\left(k^2-\frac{f''}{f}\right)\mathcal{A}(\eta,k)=0
\end{equation}

At the beginning of inflation, $\eta\approx\eta_i$.
This means $f''/f\approx 0$ and the Eq.(\ref{e35}) change to 
\begin{equation}
	\label{e51}
	\mathcal{A}''+k^2\mathcal{A}=0~~\Rightarrow~~\mathcal{A}\propto\exp(\pm i k\eta)
\end{equation}
At small scale limit, the solution should be ``negative frequency"\cite{r22}. In order to satisfy the Wronskian condition\cite{r22}, the solution of $\mathcal{A}$ should be
\begin{equation}
	\label{e52}
	\mathcal{A}=\frac{1}{\sqrt{2k}}\exp(-ik\eta)
\end{equation} 

At the late time of inflation when $\eta\approx\eta_{\infty}$ we have
\begin{equation}
	\label{e55}
	\frac{f''}{f}\approx\frac{p}{2}\left(\frac{p}{2}+1\right)T^{-2}
\end{equation}
where $T\equiv|\eta_{\infty}|-|\eta|=\eta-\eta_{\infty}$, therefore $T<0$ during the slow-roll inflation.
Notice that $dT=d \eta$, then the Eq.(\ref{e35}) change to 
\begin{equation}
	\label{e56}
	\frac{d^2}{dT^2}\mathcal{A}+\left[k^2-\frac{p}{2}\left(\frac{p}{2}+1\right)T^{-2}\right]\mathcal{A}=0
\end{equation}
The general solution of Eq.(\ref{e56}) is 
\begin{equation}
	\label{e57}
	\mathcal{A}=(-kT)^{1/2}\left[C_1(k)J_{\nu}(-kT)+C_2(k)J_{-\nu}(-kT)\right]
\end{equation}
where $\nu\equiv(1+p)/2$.

When $|\eta|\gg|\eta_{\infty}|$, Eq.(\ref{e57}) should be approximated to Eq.(\ref{e52}). While, when 
$|\eta|\gg|\eta_{\infty}|$, $T\approx\eta$, therefore Eq.(\ref{e57}) just need to be approximated to 
\begin{equation}
	\label{e58}
	\mathcal{A}=\frac{1}{\sqrt{2k}}\exp(-ikT)
\end{equation}
Eq.(\ref{e58}) can be used to determine the coefficient $C_1, C_2$.
We focus on the behavior on large scale, so we take the large scale limit: $-k\eta\rightarrow0\Rightarrow-kT\rightarrow0$. After ignoring the decay term, we have:
\begin{equation}
	\label{e64}
	\mathcal{A}\approx k^{-1/2}c(\gamma)(-kT)^{1-\gamma}
\end{equation}
where
\begin{equation}
	c(\gamma)=\sqrt{\frac{\pi}{2^{3-2\gamma}}}\frac{\exp[i\pi(1+\gamma)/2]}{\Gamma(3/2-\gamma)\cos(\pi\gamma)}\label{e63}
\end{equation}
and $\gamma\equiv1+p/2>1$. Therefore the power spectrum of magnetic field is 
\begin{equation}
	\label{e65}
	\frac{d\rho_B}{dk}=\frac{1}{k}\frac{d\rho_B}{d\ln k}=\frac{c^2(\gamma)}{2\pi^2k}H^4(k\eta)^{6-2\gamma}
	\left(1-\frac{\eta_{\infty}}{\eta}\right)^{2-2\gamma}
\end{equation}
The power spectrum of electric field can also be obtain: 
\begin{equation}
	\label{e72}
	\frac{d\rho_E}{dk}=\frac{1}{k}\frac{d\rho_E}{d\ln k}\approx\frac{d^2(\gamma)}{2\pi^2k}H^4(k\eta)^{4-2\gamma}\left(1-\frac{\eta_{\infty}}{\eta}\right)^{-2\gamma}
\end{equation}
where
\begin{equation}
	d(\gamma)=\frac{\sqrt{\pi}\exp[i\pi(1+\gamma)/2]}{2^{-\gamma+1/2}\Gamma(-\gamma+1/2)\cos(\pi\gamma)}\label{e71}
\end{equation}

From Eq.(\ref{e65},\ref{e72}) we can see that the spectral index of magnetic power spectrum is $n_B=6-2\gamma$ and the spectral index of electric power spectrum is $n_E=4-2\gamma$.
Therefore, scale invariant spectrum of magnetic field can be got when $\gamma=3$ i.e. $p=4$.
When $\gamma=2$ i.e. $p=2$ one can get the scale invariant spectrum of electric field. If the magnetic field spectrum is scale invariant ($\gamma=3$), the electric field spectrum is red.

Because of $\gamma>1$, it is can be seen from Eq.(\ref{e65}) and Eq.(\ref{e72}) that the spectrum of magnetic and electric field will increase rapidly when $\eta\rightarrow\eta_{\infty}$. This can cause backreaction problem. 

However, as we discussed above, the slow-roll inflation will ended before $\eta_{\infty}$. We assume that the moment when the slow-roll ends is $\eta_{end}$, and the value of inflaton at this moment is $\phi_{end}$.
Then we can get approximately that
\begin{equation}
	\label{e75}
	\left|\frac{\eta_{\infty}}{\eta_{end}}\right|-1\approx\ln\left|\frac{\eta_{\infty}}{\eta_{end}}\right|=-\frac{\phi_{end}^2}{2p(1+\beta)}\equiv \mathcal{Y}
\end{equation}
Insert Eq.(\ref{e75}) into Eq.(\ref{e65},\ref{e72}) one can estimate the power spectrum of magnetic and electric field at the end of slow-roll inflation:
\begin{eqnarray}
	\frac{d\rho_B}{d\ln k}\big|_{end}&\approx&\frac{c^2(\gamma)}{2\pi^2}H^4_{end}\left(\frac{k}{a_{end}H_{end}}\right)^{6-2\gamma}
	\mathcal{Y}^{2-2\gamma}\label{e76}\\
	\frac{d\rho_E}{d\ln k}\big|_{end}&\approx&\frac{d^2(\gamma)}{2\pi^2}H^4_{end}\left(\frac{k}{a_{end}H_{end}}\right)
	^{4-2\gamma}\mathcal{Y}^{-2\gamma}\label{e77}
\end{eqnarray}
To avoid the backreaction problem, the energy density of electromagnetic field can not exceed the energy density of the inflaton, this require that
\begin{equation}
	\label{e78}
	\frac{d\rho_B}{d\ln k}\big|_{end}+\frac{d\rho_E}{d\ln k}\big|_{end}<\rho_{end}
\end{equation}
We focus on scale invariant spectrum of magnetic field, i.e. $\gamma=3, p=4$, then Eq.(\ref{e78}) means
\begin{equation}
	\frac{H_{end}^4}{2\pi^2}\mathcal{Y}^{-6}d^2(\gamma)\left(\frac{k}{a_0H_0}\right)^{-2}\left(\frac{a_0H_0}{a_{end}H_{end}}\right)^{-2}
	<\frac{3}{8\pi}H^2_{end}M_{pl}^2\label{e81}
\end{equation}
The ratio $(a_0H_0)/(a_{end}H_{end})$ can be estimated as\cite{r22}:
\begin{equation}
	\label{e82}
	\frac{a_0H_0}{a_{end}H_{end}}\approx1.51\times10^{-29}\frac{h}{R},~~~~~~(h\approx0.72)
\end{equation}
where $R$ depend on the reheating phase, and for simple estimate one can chose $R\approx\rho_{end}^{1/4}$ as in \cite{r22}. Insert these into Eq.(\ref{e81}) we can have
\begin{equation}
	\label{e83}
	\rho_{end}<\left(\phi_{end}M_{pl}^{1/3}\right)^8\times10^{-42}
\end{equation}
If one require the $\rho_{end}$ should be satify the requirements of nuleosynthesis ($\rho_{nuc}\approx10^{-85}M_{pl}^4$), then
\begin{equation}
	\phi_{end}>10^{-43}M_{pl}^{4/3}	\label{e84}
\end{equation}
This means that as long as the slow-roll era ends before the inflaton decays too small, the backreaction problem can be avoided.

We also can estimate the present day value of magnetic field strength simply. From Eq.(\ref{e76}) we know that 
the power spectrum of the model in this paper is amplified by the factor $\mathcal{Y}^{2-2\gamma}$ compare with the conventional model \cite{r39,r37}.
This factor is independ on the scale factor. If one assume the instant reheating, then the present day power spectrum of magnetic field is also amplified by this factor.

While, this factor is depend on the detail of the inflation model, specifically on $\phi_{end}$. 
For example, 
for scale invariant magnetic pectrum, i.e. $V(\phi)\propto \phi^4$ , this factor is $(8/\phi_{end}^2)^4$.  
In $\phi^4$ inflation model, the slow-roll era ends at $\phi_{end}\approx M_{pl}/2$ \cite{r46}, and this factor change to $(4/\pi)^4$. Therefore, the present day value of magnetic field strength is amplified by $(4/\pi)^2\approx1.6$ compare with the conventional model \cite{r37}. This means that the magnetic field on coherence scale 1Mpc today is 
\begin{equation}
	\label{e84-3}
	B_0\approx8\times10^{-10}G\left(\frac{H}{10^{-5}M_{pl}}\right)
\end{equation}
This result satisfy the lower bound of $\gamma$-ray observation $B\sim10^{-15}G$ \cite{r6,r8}.

Notice that the lower limit of $\phi_{end}$ is very small (see Eq.(\ref{e84})), then the magnitude of the $\mathcal{Y}$ factor has a large span.
Therefore, we can use $\phi_{end}$ as a tunable parameter of the model, and adjust the value of it to make the predicted magnetic field strength of the model consistent with today's observations.
\section{\label{s4}Strong coupling}
As the last part of this paper, we will discuss the problem of strong coupling  qualitatively.
From Eq.(\ref{e44}) we know that the coupling function is monotonically increasing function during the 
slow-roll of inflation. This will lead to the strong coupling problem which was first pointed out in \cite{r43}. To avoid this problem, one can assume 
a decreasing coupling function during the preheating era like in \cite{r41}. 

However, in this paper, we found that the coupling fucntion should satisfy the Eq.(\ref{e26}).
An attractive possibility is that this equation can lead to an decrease coupling function after the slow-roll of inflation.
Therefore it is interesting to discuss the 
behavior of coupling function during the preheating.

It should be noted that, Eq.(\ref{e44}) is satisfied only in slow-roll era. During the preheating, the coupling function should be obtain by solving the Eq.(\ref{e24},\ref{e26}) together. 
One can eliminate the $A_j$ by combining Eq.(\ref{e24},\ref{e26}) and get
\begin{equation}
	\label{e94}
	\mathcal{D}_1h'+\mathcal{D}_2h+\mathcal{D}_3h^2+\mathcal{D}_4=0
\end{equation}
where 
\begin{eqnarray}
	&&\mathcal{D}_1=\phi'^2-2\mathcal{H}q\\
	&&\mathcal{D}_2=4q\phi'^2-8\mathcal{H}q^2-2\phi'\phi''+2\mathcal{H}'q+2\mathcal{H}q'\\
	&&\mathcal{D}_3=-\phi'^2+2\mathcal{H}q\\
	&&\mathcal{D}_4=-4\phi'^2q'-\phi'^4+4\mathcal{H}q\phi'^2-4\mathcal{H}^2q^2\nonumber\\
	&&~~~~~~+8\phi'\phi''q-8\mathcal{H}'q^2\\
	&&q\equiv \frac{\Phi}{\delta\phi}\phi',~~h\equiv \mathcal{G}\phi'=2\frac{f'}{f}
\end{eqnarray}
Eq.(\ref{e94}) is the equation that the coupling function needs to satisfy during the preheating. On large scale, the Fourier mode of Bardeen potential and perturbation of inflaton can be written as \cite{r47}
\begin{equation}
	\label{e95}
	\Phi_k=\mathcal{C}\left(1-\frac{H}{a}\int adt\right),~~~\delta\phi_k\approx\mathcal{C}\dot{\phi}\left(a^{-1}\int adt\right)
\end{equation}
Consider the case where the magnetic field is scale invarant $p=4$, which means that we should consider the 
$\phi^4$ preheating. In this model of preheating, $a\propto\sqrt{t}$ and the evolution of $\phi$ can be approximated as \cite{r46}
\begin{equation}
	\label{e97}
	\phi\approx\frac{\tilde{\phi}}{a}\cos(0.8472\sqrt{\lambda}\tilde{\phi}\eta)
\end{equation}
For qualitative discussion, we assume
\begin{equation}
	\label{e98}
	\phi=\frac{1}{\eta}\cos\eta
\end{equation}

\begin{figure}
	\includegraphics[width=0.8\columnwidth]{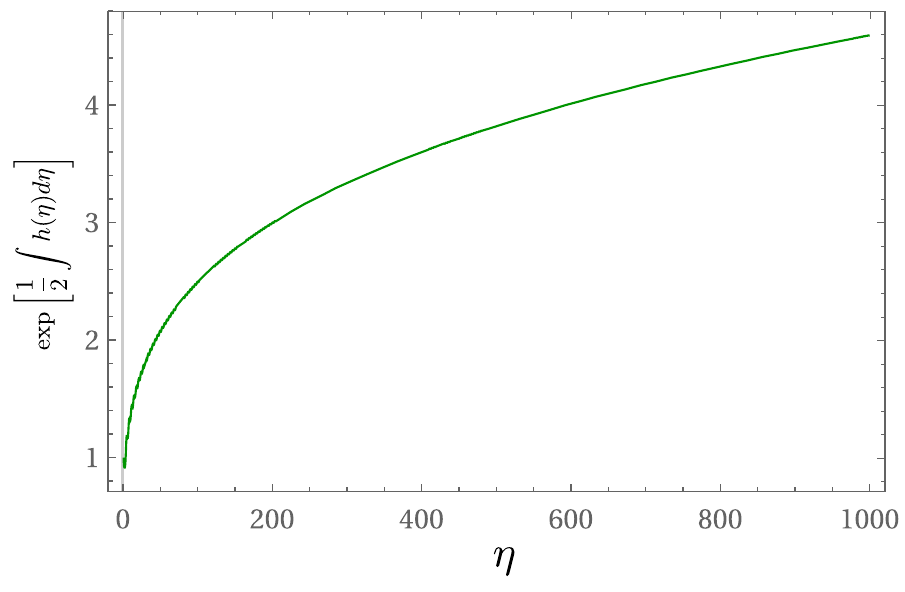}
	\caption{\label{f1}Coupling function during $\phi^4$ preheating with no correction.}
\end{figure}
The evolution curve of the coupling function given in the Fig.\ref{f1}. It can be seen that the coupling function will increase during the preheating era. This will lead to strong coupling problem. 

\textcolor{black}{
	In conventional models, one strategy to solve strong coupling problem is to ansatz a decreasing coupling function during the preheating era directly as in \cite{r41}. However, in the models we consider here, the coupling function should satisfy Eq.\eqref{e94}, which means that the coupling function is model dependent. In other words, to get the decreasing coupling function, we should modify the preheating model.
}

\textcolor{black}
{
We consider a toy model in which there are some modifications at the beginning of $\phi^4$ preheating. As the preheating process goes on, the modificatons will vanish and the preheating model will approximate to the $\phi^4$ model. 
}

\textcolor{black}{
These modifications affect both the dynamics of background $\phi,~\mathcal{H}...$ and perturbation $\Phi,~\delta\phi...$, and affect the Eq.\eqref{e94}. We assume that the effect of  these modifications of preheating model change the Eq.\eqref{e94} to 
\begin{equation}
	\label{e94-1}
	\mathcal{D}_1\hat{h}'+\mathcal{D}_2\hat{h}+\mathcal{D}_3\hat{h}^2+\mathcal{D}_4=0
\end{equation}
where 
\begin{equation}
	\label{e100}
	\hat{h}(\eta)\equiv h(\eta)+r(\eta)
\end{equation}
and $r(\eta)$ is a decreasing function of $\eta$. In Eq.\eqref{e94-1}, the dynamics of $\phi,\mathcal{H},\delta\phi,\Phi$ are the same as in $\phi^4$ model, which means that we assume the modifications of preheating model equivalent to introduce a correction function $r(\eta)$ in Eq.\eqref{e94}.
}

\textcolor{black}{
As the preheating proceeds, the correction function will tend to zero
and $\hat{h}\rightarrow h$, therefore $h$ will satisfy the Eq.(\ref{e94}) at late time of preheating and the preheating model approximate to the $\phi^4$ model as we assumed.
}

\textcolor{black}{
One simple choice of $r(\eta)$ is 
\begin{equation}
	\label{e101}
	r(\eta)=\frac{1}{\eta^n}
\end{equation}
Notice that the form of $r(\eta)$ depend on the  modifications of preheating model, then the parameter $n$ can be seen as a parameter of  preheating model. This parameter can be choosen to avoid the strong problem. 
}

\begin{figure}
	\includegraphics[width=0.8\columnwidth]{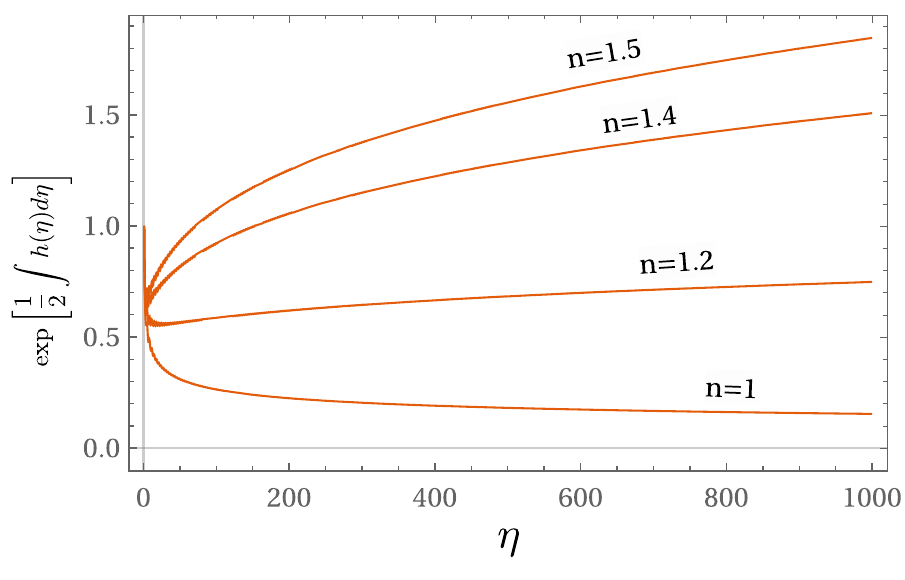}
	\caption{\label{f2}Coupling function during $\phi^4$ preheating with different correction functions.}
\end{figure}
The evolution of coupling function during preheating with different choices of $r(\eta)$ are given in Fig.\ref{f2}.
It can be seen that a decreasing coupling function can be obtained by selecting the appropriate parameter n (i.g. $n=1$ in Fig.\ref{f2}), and then the strong coupling can be avoided.

\section{\label{s5}Summary and discussion}
In this paper, we discuss the inflationary magnetogenesis with a coupling function which can keep the action to be self-consistent. This self-consistence coming from the time component of Maxwell's equation which is the secondary constraint for electromagnetic field. Under the FRW metric, this is a trival equation. However, once we consider the perturbed metric, this equation become non-trival and can be seen as a 
restrict equation for coupling function $f(\phi)$, see Eq.(\ref{e21}). Taking this as a starting point, we calculated the power spectrum of the electric and magnetic fields in the large-scale inflation model.
We estimate the present day value of magnetic field, and the result satisfy the lower bound of $\gamma$-ray observation.

We found that, the power spectrum obtained in this  paper is multiplied by a factor related to $\phi_{end}$ compared to the conventional  model\cite{r39,r37}. This means that one can generate the required magnetic field by choosing a suitable inflation model, or conversely use the magnetic field observed today to limit $\phi_{end}$. This provides a possibility to use today's large-scale magnetic field strength to estimate the value of the inflaton field at the end of the slow-roll.

On the other hand, to avoid the backreaction problem at the end of inflation, the value of inflaton at the end of slow-roll era $\phi_{end}$ should have a lower limit ($\sim10^{-43}$)(see Eq.(\ref{e84})). This lower limit is so small that the upper limit of the factor $\mathcal{Y}$ can be large. This means that the model discussed in this paper can generate a sufficiently strong magnetic field without causing backreaction problem.

The strong coupling problem can also appear in the situation which is discussed here. One way to solve the problem of strong coupling is to introduce an decreasing coupling function in the preheating era. In this paper, the coupling function is determined by Eq.(\ref{e8}) or Eq.(\ref{e94}) in preheating era. Unfortunately these equations give an increasing coupling function. To make the coupling function change to a decreasing function, it is necessary to introduce a correction function in the early stage of preheating.
So  it is a very interesting open topic to find a suitable preheating model in which the coupling function can naturally be determined as an decreasing function.

\section*{Acknowledgments}
This work was supported by the Fundamental Research Funds for the Central Universities of
Ministry of Education of China under Grants No. 3132018242,
the Natural Science Foundation of Liaoning Province of China under Grant No.20170520161 and the National Natural Science Foundation of China under Grant No.11447198 (Fund of theoretical physics).
\bibliographystyle{ws-mpla}
\bibliography{leeyu-mpla}
\end{document}